# Solution-corrected Constant Potential Model for CO$_2$ Electrocatalysis in Ionic Liquids

*Jikai Sun[1, \*], Alejandro Gallegos[2], Runtong Pan[1], and Jianzhong Wu[1,\*]*

[1]Department of Chemical and Environmental Engineering, University of California, Riverside, CA 92521, USA

[2]Department of Chemical & Materials Engineering, New Mexico State University, Las Cruces NM 88003, USA

**Abstract**

The selection of suitable ionic liquids (ILs) is critical for CO$_2$ capture and electrocatalytic conversion into valuable chemical products. The screening process can be enhanced with theoretical simulations that reveal the property-performance relationship of ILs, accelerating the identification of optimal candidates. However, anhydrous ionic liquids exhibit low dielectric constants and high ion concentrations, challenging traditional first-principles calculations. Additionally, the spatial distribution of CO$_2$ in the electric double layer (EDL) plays a crucial role in determining the electrocatalytic activity. This work proposes a solution-corrected constant potential model (CPM-sol) to account for the imbalance between the net charge and the number of electrons on the electrode surface through an implicit consideration of ion distributions. By incorporating solution-phase corrections into the conventional CPM model, we reveal the changes in the Fermi level and charge alongside the reaction process. Furthermore, we systematically investigate the impact of various IL properties on electrode surface charging and CO$_2$ distribution. The theoretical results highlight the critical role of interactions between solution components, forming chain-like structures, in determining their distribution in confined environments and influencing the electrode surface charge. These findings provide insights for mechanism-guided electrolyte design.

**Keywords**

*To whom correspondence should be addressed. Email: jikais@ucr.edu; jianzhong.wu@ucr.edu

solution-corrected constant potential model (CPM), ionic liquids, electric double layer (EDL), $CO_2$ distribution, confinement conditions.

**Introduction**

Electrocatalytic $CO_2$ reduction reaction ($CO_2$RR) is of great significance in renewable energy conservation and greenhouse gas emission control.[1] Among various strategies, a promising approach involves using anhydrous ionic liquids (ILs) as solvents and immersing them into porous materials for $CO_2$ capture and conversion.[2–5] Their high $CO_2$ capture capacity, along with the ability to suppress side reactions such as hydrogen evolution reactions (HER), makes ILs highly promising for $CO_2$RR applications.[6,7] Understanding the relationship between various properties of ionic liquids and their activity in $CO_2$ conversion is crucial for guiding the selection and regulation of best candidates for targeted applications. Key factors influencing their catalytic performance include cation-anion interactions, their respective affinities with $CO_2$, and interactions of all chemical species with the electrode surface. These properties collectively determine the efficiency of $CO_2$ capture and the stability of reaction intermediates.[8,9] Experimental screening of ILs, while valuable, is not only time-consuming but also labor-intensive. Furthermore, these properties are often interdependent, meaning that modifications to one parameter can lead to cascading changes in others.[10] Therefore, assessing the influence of a single property on ionic liquid performance through experimental approaches alone presents a significant challenge.

First-principles calculations provide atomic-level insights into electrocatalytic mechanisms.[11,12] The constant charge model (CCM), one of the most widely used frameworks, effectively captures key aspects of the electrocatalytic process under a fixed electrode charge.[13] In recent years, an increasing number of studies have adopted the constant potential model (CPM), which better aligns with experimental conditions.[14–16] Unlike CCM, which maintains a fixed number of electrons on the electrode surface, CPM assumes that the electrode potential remains constant during electrochemical reactions. Despite their usefulness, neither model adequately accounts for the influence of the solution environment from a molecular perspective. In particular, the first-principles methods often rely on simplified models for describing the electric double layer (EDL) at the electrode surface, neglecting ion-specific effects on electrocatalytic behavior. Because different electrolytes generate distinct interfacial charges under the same electrode potential, they consequently affect the electron density on the electrode surface and alter the reaction process. This variability disrupts the charge compatibility within EDLs due to the spatiotemporal mismatch between electron transfer and ion diffusion. As a result, interfacial electrons are redistributed to neutralize localized electric fields and supplement the surface charge, influencing electrocatalytic performance and reaction kinetics.[17–19]

Owing to the extremely high ion concentration and low permeability of ionic liquids, the EDL thickness tends to be much smaller than that in an aqueous solution, resulting in a more concentrated potential drop across the electrode -electrolyte interface. This pronounced effect significantly influences interfacial charge distribution and electrostatic interactions, making it a crucial factor in electrocatalytic processes.[20] In addition, the distribution of $CO_2$ within the ionic liquid, especially its accumulation at the electrode surface, plays a crucial role in determining the reaction kinetics.[20–23] It was reported that the reactivity of $CO_2$ in ionic liquids differs from its solubility order.[2,8,24] On one hand, this discrepancy arises from different interactions between ionic liquids and the electrode surface, variations in reaction intermediates, and other factors, leading to differences in surface electrode potential and intermediate stability.[1,2,25,26] On the other hand, this reflects the fact that $CO_2$ solubility in the bulk phase does not necessarily correlate with its concentration at the surface, which requires a more precise assessment of its localization and reactive behavior.

To address the aforementioned theoretical challenges, namely the impact of EDL on the electrode surface charge and the distribution of reactant molecules in the electrolyte, we need to explicitly consider the molecular characteristics of the local solution environment.[27–29] However, incorporating explicit solution molecules into first-principles simulation greatly increases the computational burden, making it difficult to balance between accuracy and efficiency. Moreover, conventional molecular simulations often rely on closed systems with a fixed number of molecules. The canonical ensemble cannot fully capture the open nature of experimental environments, thereby limiting the applicability of first-principles models. In contrast, classical DFT (cDFT) operates under the grand canonical ($\mu VT$) ensemble, allowing for fluctuations in particle numbers.[30,31] By minimizing the grand potential, cDFT enables the determination of the spatial distribution of solution components under equilibrium conditions, making it a powerful tool for modeling electrochemical interfaces.[32,33] Importantly, cDFT does not require sampling of molecular motion trajectories and directly outputs the equilibrium density distributions. This feature makes cDFT particularly well-suited for studying interfacial phenomena, where capturing the spatial organization of solution components is essential.

In this work, we present a hybrid computational framework that integrates cDFT with the Kohn–Sham density functional theory (KS-DFT). Within this approach, cDFT is employed to model the solution environment beyond the conventional CPM model. This setup enables the self-consistent generation of the EDL structure, which in turn influences and adjusts the electron density on the electrode surface through feedback into the KS-DFT calculations. We designate this hybrid DFT approach as the solution-corrected constant potential model (CPM-sol). This approach not only captures the impact of the solution environment on the electrode potential and electron density but also provides detailed spatial distribution of solution species. As a result, it reveals key interfacial phenomena such as the enrichment or exclusion of reactants and $CO_2$ molecules at the electrode surface. Leveraging the enhanced capabilities of the CPM-sol model, we systematically investigate the distribution of $CO_2$ molecules in ionic liquids confined within micropores of model electrodes. We explore how various properties of the electrode surface and ionic species affect the EDL structure and surface energy. Additionally, we characterize the variation patterns of the EDL charge under different conditions. Finally, we apply the CPM-sol model to simulate the C-C coupling process on the Cu (100) surface, uncovering mechanistic insights that extend beyond those predicted by the traditional CPM model.

**Method**

*CPM-sol Framework*

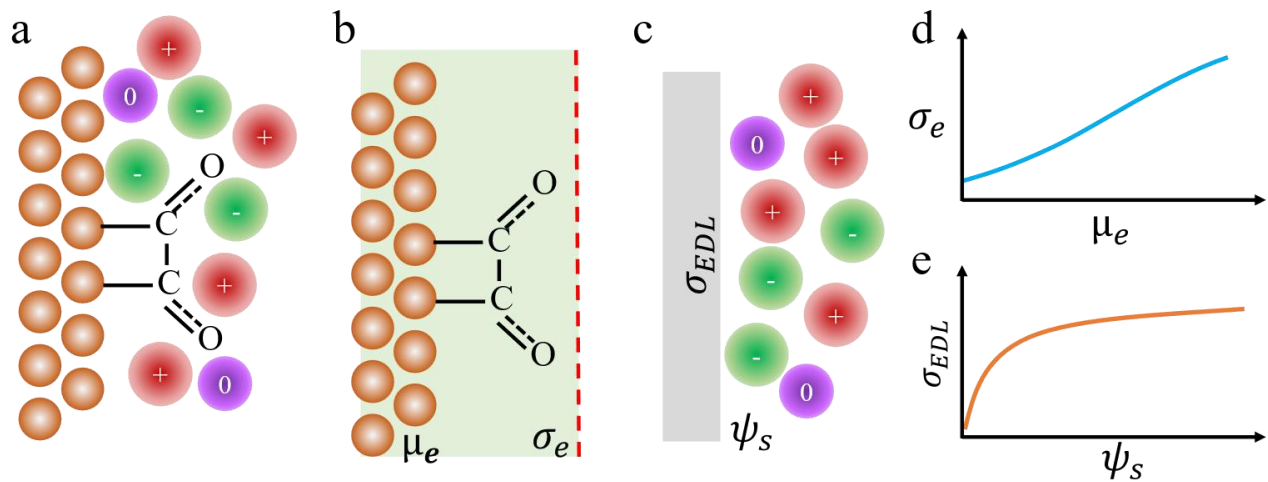

Figure 1 Schematic of the solution-corrected constant potential model (CPM-sol) for electrocatalyst in ionic liquids: (a) an intermediate state (e.g., *CCOO) surrounded by reactants and ionic species near the electrode surface; (b) the system setup for KS-DFT calculations to establish the surface charge density ($\sigma_e$) and Fermi energy ($\mu_e$) relationship in a dielectric medium, as shown in panel (d). (c) the electric double layer (EDL) described by cDFT, which predicts the EDL charge density ($\sigma_{EDL}$) as a function of the surface potential ($\psi_s$). While the charge neutrality is satisfied in both KS-DFT and cDFT calculations, $\sigma_e \neq \sigma_{EDL}$ due to approximations in both quantum and classical models. The red, green, and purple spheres represent cations, anions, and neutral reactants (e.g., $CO_2$), respectively.

Figure 1 illustrates schematically key electrochemical variables that will be utilized to analyze the effects of ionic liquids on electrocatalytic $CO_2$ reduction reaction ($CO_2$RR). For simplicity, our analysis focuses on the thermodynamic properties of CO–CO intermediate on the Cu (100) surface as a benchmark. This surface site plays a crucial role in C–C coupling for electrocatalytic $CO_2$RR toward $C_2^+$ products.[26,34] For the electrochemical reaction in an ionic liquid, we assume that the cations and anions, as well as the $CO_2$ molecules, directly interact with the electrode surface, without the presence of a Stern layer.[24,35]

The work function of the electrode changes with the electrical potential, thereby regulating the surface charge and electronic structure.[14,36,37] In the traditional constant-potential model (CPM), the relationship between the surface charge density ($\sigma_e$) and the electrode potential is established through KS-DFT calculations under a fixed dielectric constant. In other words, CPM assumes a static EDL and its response to changes in the electrode potential, which is controlled by the Fermi level, is accounted for by an effective surface charge ($\sigma_e$). Under a given electrode potential, an electric double layer (EDL) is formed due to the rearrangement of ions and various chemical species in the electrolyte solution, resulting in a net charge ($\sigma_{EDL}$) that counterbalances the applied potential at the electrode. While the electrostatic neutrality requires that the electrode charge ($\sigma_e$) must balance the net charge of ionic species in the EDL ($\sigma_{EDL}$), these charge densities may not be the same when they are predicted from different theoretical models. In this work, $\sigma_e$ is obtained from KS-DFT calculations, maintaining a constant Fermi level, while $\sigma_{EDL}$ is simulated using classical DFT. These simulation methods are independent to each other and reasonable only within their respective assumptions. On one hand, KS-DFT captures the changes in the electrode charge under different surface and adsorption conditions at a fixed electron chemical potential or electrode potential. On the other hand, cDFT enables the simulation of the ion distributions in the solution phase across a range of electrolyte types and concentrations, also under a given electrode potential. In other words, KS-DFT describes how the applied voltage shifts the electrochemical potential of the electrode, as reflected in the Fermi level, promoting the electrode to adjust its electron content in response. Conversely, cDFT simulates how ions in the surrounding solution responds to the electrode potential by redistributing through adsorption or diffusion, thereby forming countercharges.

However, the electrode surface charge density obtained from the conventional CPM calculations is usually different from the EDL charge predicted via cDFT. This discrepancy arises because KS-DFT, as used in CPM, typically neglects the influence of the solution environment when calculating the electrode's Fermi level. In contrast, cDFT simulates the solution environment often based on an idealized representation of the electrode as a charged hard wall. Although the charge neutrality is maintained in both KS-DFT and cDFT calculations, these idealized assumptions lead to differences in the electrode surface charge densities obtained by KS-DFT and cDFT, thereby creating a charge imbalance at the interface. Nevertheless, due to the independent implementation of these two approaches, such inconsistencies are often overlooked. That is, most existing methods rely on the initially established relationship between potential and charge, without accounting for the emergence of a residual net interfacial field and the subsequent rebalancing process. In reality, when net charge accumulates at the interface, it generates an additional electric field that perturbs both the electronic distribution within the electrode and electrolyte distribution within the solution. To restore electrochemical equilibrium, the electron population in the electrode and the spatial distribution of ions in the electrolyte must adjust accordingly. The system reaches a true equilibrium state when the Fermi level aligns with the electrochemical potential required to balance the interfacial charge between the electrode and the electrolyte.

For instance, if the charge density at the electrode surface ($\sigma_e$) exceeds that caused by the net charge of ion species in the EDL ($\sigma_{EDL}$), the ionic solution will increase the accumulation of cations at the surface, thereby increasing $\sigma_{EDL}$. While for the electrode electrons, their arrangement in response to the EDL charge tends to reduce the number of electrons. Ultimately, the true electrode surface charge density ($\sigma_{act}$) should lie between these two values. The specific variation is a complex iterative process, and in this work, we approximate it as the average, $\sigma_{act} = (\sigma_e + \sigma_{EDL})/2$. Although this treatment is admittedly coarse, it qualitatively captures the competing effects of Fermi level modulation and ionic redistribution in the electrolyte. It is important to emphasize that the consideration of the solution leads to a change in the number of electrons on the electrode surface, which also causes the Fermi level to shift. Therefore, the relationship between the Fermi level and electrode potential is no longer a simple linear relationship. Additionally, the relationship between the number of electrons and Fermi level varies with different electrodes or electrodes with different adsorbates. Even for the same solution, different charge differences can arise due to adsorption, leading to different Fermi levels. In other words, for a specific solution and the reaction on the electrode surface, the Fermi level will vary along with the reaction process. This behavior undoubtedly reflects the realistic nature of electrochemical interfaces more accurately.

Here, we emphasize that the CPM-sol model builds upon the conventional CPM framework by incorporating cDFT to simulate the spatial distribution of electrolyte species. Through this coupling, the CPM-sol model accounts for the influence of the electrolyte environment on the surface potential and interfacial charge of the electrode, effectively integrating solution effects into the CPM approach. This represents a significant advancement over the traditional CPM model, which largely neglects the EDL effects, and thus offers a more physically consistent description of electrified interfaces.

Nevertheless, it should be recognized that the CPM-sol model inherently relies on an implicit representation of the solution environment. In situations where chemical adsorption of solvent or electrolyte species occurs, or when these species directly participate in the reaction, explicit modeling -- such as that enabled by KS-DFT or ab initio molecular dynamics (AIMD) -- becomes indispensable. In such cases, the electrostatic corrections to potential and charge proposed in this study can still serve as a valuable reference, guiding and improving the accuracy of explicit simulations.

**Results and Discussion**

*Application of CPM-sol Framework*

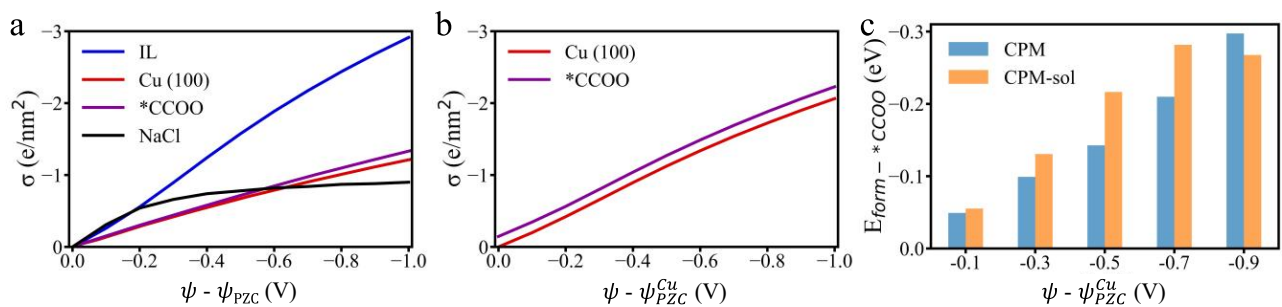

Figure 2 Variation of the electrode surface charge density and the *CCOO formation energy with the electrode potential: (a) the electrode surface charge density as a function of the electrode potential for pristine Cu (100), *CCOO-adsorbed Cu (100), ionic liquid (IL), and 1 M NaCl solutions; (b) the electrode surface charge density versus the electrode potential for the pristine and *CCOO-adsorbed Cu (100) after correction by the ionic liquid environment; (c) the formation energy of *CCOO as a function of electrode potential under the conventional constant potential model (CPM) and the IL solution-corrected model (CPM-sol). Note that in panel (a), the electrode potentials are referenced to the respective potential of zero charge (PZC) for each condition, with different PZC values for pristine Cu (100) and *CCOO-adsorbed Cu (100), while in panels (b) and (c), the electrode potentials are referenced to the PZC of pristine Cu (100).

We first compare the theoretical predictions of CPM and CPM-sol models for the $CO_2$ coupling reaction on Cu (100) surface, in which the stability of *CCOO is considered as the key points of the reaction activity.[34] Figure 2a shows the variation in the electrode surface charge density with electrode potential for pristine Cu (100) and *CCOO-adsorbed Cu (100) surfaces, which were obtained by KS-DFT calculations. It is important to note that both the pristine Cu (100) and *CCOO-adsorbed Cu (100) surfaces are referenced to their respective potential of zero charge (PZC). As shown in Figure S1, their PZCs are different. From Figure 2a, it can be observed that even when referenced to their respective PZCs, the electrode surface charge density curves for pristine Cu (100) and *CCOO-adsorbed Cu (100) are not completely identical. The difference gradually increases after the electrode potential exceeds -0.6 V. Furthermore, Figure 2a presents the surface charge density versus the electrode potential resulting from the ionic rearrangement in the ionic liquid and in a 1 M NaCl aqueous solution. These results were simulated using classical DFT as detailed in Supporting Information (SI). For both the ionic liquid and 1M NaCl solutions, the EDL charge density ($\sigma_{EDL}$) increases at a similar rate under low electrode potentials, significantly faster than the increase in electron density ($\sigma_e$). As the electrode potential increases, the charge density of 1M NaCl solution gradually approaches saturation, resulting in a charge density lower than $\sigma_e$. However, for the ionic liquid, the abundance of cations and anions, along with the anhydrous shielding effect, causes the electrode charge density to continue to increase with the electrode potential, significantly surpassing $\sigma_e$.

It is important to note that for any electrode surface, its PZC varies with the solution condition. At the PZC, both the surface charge density of the electrode ($\sigma_e$) and the charge density generated by ion rearrangement ($\sigma_{EDL}$) are zero. While different electrode-solution combinations have different PZCs, the electrode and solution in each combination have a single PZC. Based on the combination of KS-DFT and classical DFT calculations, we obtained the growth curves of solution-corrected electrode surface charge density ($\sigma_{act}$) with electrode potential for both pristine Cu (100) and *CCOO-adsorbed Cu (100) surfaces in an ionic liquid environment, as shown in Figure 2b. The high charge density-electrode potential response of the ionic liquid greatly increases the number of electrons on the Cu surface at a given electrode potential. It should be noted that the solution-interface charge difference ($\sigma_e \neq \sigma_{EDL}$) leads to changes in the surface charge density of the electrode, which also causes corresponding changes in their Fermi levels. Since the $\sigma_e$ of pristine Cu (100) and *CCOO-adsorbed Cu (100) surfaces are different, their Fermi levels are no longer the same!

Figure 2c shows the formation energy of *CCOO ($E_{form-*CCOO}$) obtained using the traditional CPM model and our solution-corrected CPM-sol model. All the energies were referred to the $E_{form-*CCOO}$ at 0.0 V vs PZC of pristine Cu (100) surface ($\psi_{PZC}^{Cu}$). The CPM model predicts that the *CCOO formation energy increases monotonically with the increasing electrode potential. The CPM-sol model, which considers the correction from the ionic liquid, results in a more negative surface charge at the same electrode potential, leading to a lower *CCOO formation energy. At the same time, the CPM-sol model reduces the charge density difference between pristine Cu (100) and *CCOO-adsorbed Cu (100) surfaces to some extent. And, more negative surface charges correspond to higher Fermi levels, as a result, the $E_{form-*CCOO}$ calculated by the CPM-sol model decreases at higher negative electrode potentials (-0.9 V vs $\psi_{PZC}^{Cu}$). This indicates that simply increasing the applied voltage may actually be detrimental to the reaction, reducing both reaction efficiency and selectivity.[38–40]

### *Effect of Ionic Liquid Properties on Surface Charge Density*

The choice of ionic liquid has a significant impact on the C-C coupling activity. However, the underlying mechanisms of how the ionic liquid properties affect the reaction activity are still not fully understood. This section provides a systematic investigation of various properties of ionic liquids to reveal their influence on the electrode surface charge density. First, we explore the effect of cation-anion interactions beyond electrostatic forces, as represented by the energy parameter in the square-well potential (denoted as $\varepsilon_{ion}$). The additional potential may arise from van der Waals attraction, hydrogen bonding, and electron donor-acceptor effects. Figure 3a presents the influence of $\varepsilon_{ion}$ on the electrode surface charge density. The repulsive forces between the cations and anions increase the value of $\sigma$, while the increase in the attractive forces between the cations and anions causes $\sigma$ to gradually decrease to zero. This is because strong cation-anion attractions lead to aggregation, which reduces the ion density in the solution, as explored in the next section. Figure 3b shows the value of $\sigma$ as a function of $CO_2$ solubility. As the molar fraction of $CO_2$ increases, $\sigma$ gradually decreases. Figure 3c reveals the influence of the interaction between cations/anions and $CO_2$ ($\varepsilon_{cation-CO2}/\varepsilon_{anion-CO2}$) on $\sigma$, displaying a volcano-type relationship. When the interaction between cations/anions and $CO_2$ is around 5 $k_BT$, the electrode surface charge density reaches its minimum.

Additionally, we investigate the dependence of $\sigma$ on the electrode surface interaction with cations/anions ($\varepsilon_{s-ion}$) and with $CO_2$ ($\varepsilon_{s-CO2}$), which varies the properties of the ionic liquid and the specific electrode surface. As shown in Figure 3d, a stronger attraction between the electrode surface and cations/anions results in a lower $\sigma$ value. At low surface charge densities, the densities of cations and anions at the surface are low, and the electrode's attraction to $CO_2$ has little effect on the $\sigma$ value. However, when the surface charge density is low, a strong $\varepsilon_{s-CO2}$ will attract more $CO_2$ molecules, displacing the cations and anions on the surface, which leads to a decrease in σ. Finally, Figure 3e illustrates the impact of the ionic liquid cation-anion diameters on the electrode potential, which primarily depends on the cation diameter. The larger the cation diameter, the lower the surface density, which leads to a lower surface charge density.

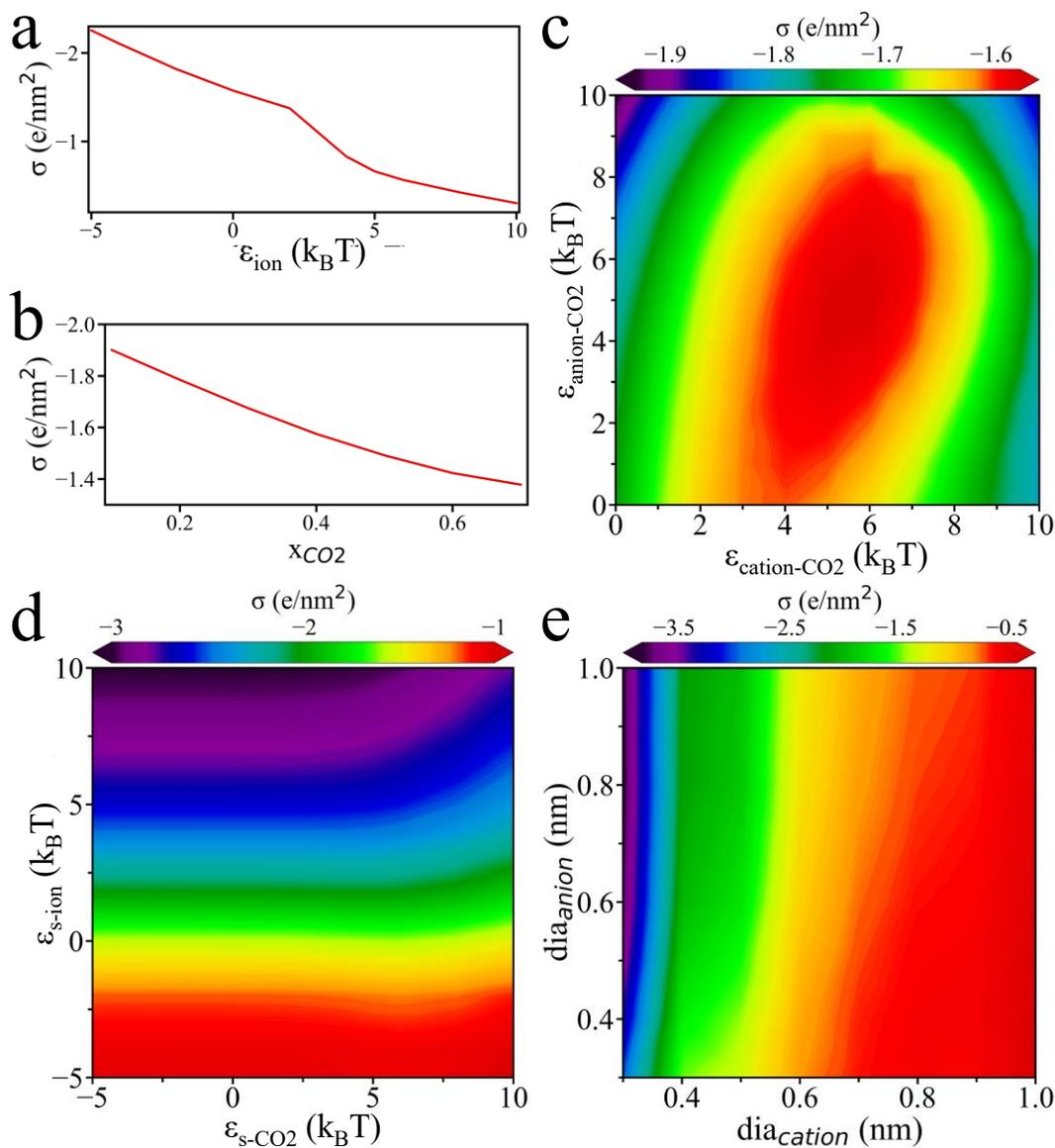

Figure 3 Variation of the electrode surface charge with IL properties: (a) dependence of electrode surface charge on the interaction strength between IL cations and anions ($\varepsilon_{ion}$); (b) variation of the electrode surface charge with the molar fraction of dissolved $CO_2$ ($x_{CO2}$); (c) 2D heatmap showing the effect of IL cation/anion–$CO_2$ interactions ($\varepsilon_{cation-CO2}/\varepsilon_{anion-CO2}$) on electrode surface charge; (d) 2D heatmap showing the influence of electrode-ion and electrode-$CO_2$ interactions ($\varepsilon_{s-ion}$, $\varepsilon_{s-CO2}$) on surface charge; (e) 2D heatmap illustrating the dependence of electrode surface charge on the diameters of IL cations and anions.

*Mechanism of Ionic Liquid Distribution*

To explain the intrinsic mechanism behind the effect of ionic liquids on the electrode surface charge density, and to visualize the distribution of $CO_2$ in ionic liquids, we conducted a detailed analysis of the density profiles of ionic liquid components. Figures 4a-c show the influence of ion-ion interaction energies on the distribution of various components under an electrode potential of -0.5 V vs PZC. The ion-ion interaction energy ranges from -5 $k_BT$ (repulsive) to 5 $k_BT$ (attractive) to 10 $k_BT$ (strong attraction), which influences the arrangement of ionic liquid components. As shown in Figure 4a, a repulsive interaction between the cations and anions leads to long-range oscillations to gradually neutralize the effect of the electrode potential. However, when the cation-anion interaction is attractive, the effect of the electrode potential can be quickly neutralized without oscillation. But at positions far from the electrode potential, the ion concentrations ions are lower than their bulk values. At 10 $k_BT$ strong attraction, the anion concentration decreases to zero. Figures 4d-f display the same decrease in the concentration of IL components at -0.0 V vs PZC under strong attractive forces. Figure 4f shows that at 10 $k_BT$, the cations and anions appear approximately 9 nm away from the surface.

The analysis suggests that the attraction between cations and anions leads to the formation of long -cation-anion-cation-anion- chains, as shown in Figure 4g. Due to various attractive forces, long chains of cations and anions, as well as their binding with $CO_2$ molecules, are formed in the solution. When the chain length exceeds the slit width, the chain cannot enter the slit pore. As shown in Figure 4h, the distribution of chain lengths in relation to the cation-anion attraction energy can be approximated by a normal distribution; the stronger the attraction, the longer and more numerous the chains. Therefore, for different $\varepsilon_{ion}$ values, the density of various components at the middle of the two hard walls ($\rho_{mid}$) as a function of slit width is shown in Figure 4i. Strong attractive forces combined with slit confinement lead to a density much lower than the bulk density. Thus, in Figure 4f, the cations and anions begin to appear at 9 nm from the surface and gradually increase, indicating that the strong attraction of 10 $k_BT$ causes the length of the -cation-anion- chain to exceed 9 nm.

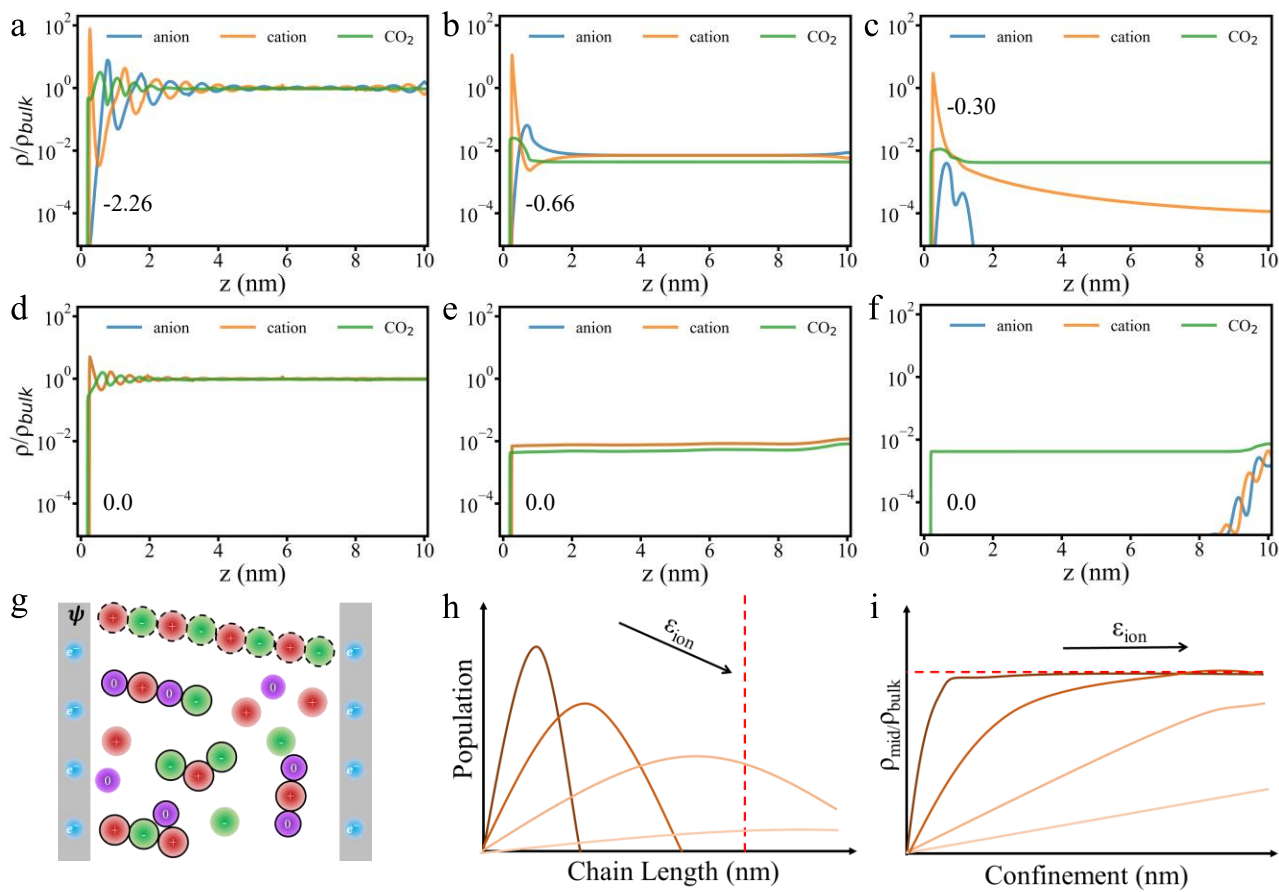

Figure 4 Density profiles under different strengths of cation–anion interaction. Density profiles at -0.5 V (vs. PZC) with $\varepsilon_{ion}/k_BT$= –5 (a), 5 (b), and 10 (c); and density profiles at 0.0 V (vs. PZC) with $\varepsilon_{ion}/k_BT$= –5 (d), 5 (e), and 10 (f). The surface charge densities were marked. g. Scheme of solution components combining into chain-like structures under the influence of interaction forces. h. Schematic illustration of the population of chain lengths of chain-like structures under different ion interaction strengths. i. Schematic illustration of the ratio of the density at the middle of two hard walls in a confined environment to the bulk density, as a function of slit width, under different ion interactions.

The curve showing the variation of ionic liquid components with the molar fraction of dissolved $CO_2$ further supports our proposed mechanism. As shown in Figures S2 and S3, when the $CO_2$ molar fraction exceeds 0.6, the concentration of IL components far from the surface starts to drop below their bulk density. This is because the default interaction between the cations/anions and $CO_2$ is 5 $k_BT$. When the $CO_2$ molar fraction is low, there are not enough $CO_2$ molecules to form -cation-$CO_2$-anion-$CO_2$- long chains. When the $CO_2$ molar fraction reaches 0.6 or 0.7, the long chains formed by sufficient $CO_2$ significantly lower the concentration of IL components in the slit. When the attraction between cations/anions and $CO_2$ is zero, as shown in Figure S4, large $CO_2$ molar fractions no longer affect their concentration far from the surface. Moreover, the increase in $CO_2$ molar concentration no longer affects the surface charge density. This indicates that the impact of $CO_2$ concentration primarily arises from its interaction with the cations and anions, which alters the size of the cations and anions. When there is no interaction between the cations/anions and $CO_2$, although the increase in $CO_2$ molar fraction reduces the concentration of the ionic liquid, there are still sufficient cations and anions accumulating on the surface, which does not affect the double-layer structure and surface charge density.

It is generally believed that the high interaction energy of ionic liquids with $CO_2$ leads to their high solubility. Previous studies have shown that anions play the main role in tuning $CO_2$ capture, but the partial negative charge on the oxygens of $CO_2$ may interact with cation centers.[7,41,42] To further illustrate this effect, we separately explored the variation of the interaction energy between the cations/anions and $CO_2$ on the local concentration of $CO_2$ molecule near the electrode surface, as shown in Figures 5 and S5. Figure 5a shows that when there is no interaction between the cations/anions and $CO_2$, $CO_2$ appears closer to the electrode surface due to its small volume. However, due to the accumulation of cations and anions at the surface, $CO_2$ is pushed away and its density near the surface becomes lower than its bulk value. Figures 5a-c indicate that when the anion-$CO_2$ interaction ($\varepsilon_{anion-CO2}$) is zero, and the cation-$CO_2$ interaction ($\varepsilon_{cation-CO2}$) changes from 0 to 5 $k_BT$ of attraction, cation-$CO_2$ chain-like structures form at the surface, significantly increasing $CO_2$ concentration at the cation locations. At the same time, this weakens the concentration of cations, which results in a decrease in surface charge density. When the $\varepsilon_{cation-CO2}$ interaction is stronger (10 kBT), more cation-$CO_2$ pairs form, significantly increasing $CO_2$ surface concentration. However, this also reduces the cation concentration, making it harder for the cations at the surface to counterbalance the effect of the electrode potential. This causes oscillations in the IL composition with distance, leading to an anomalous increase in net charge in the solution. Figures 5a, 5d, and 5g show the same trend when the $\varepsilon_{cation-CO2}$ is zero, and the $\varepsilon_{anion-CO2}$ increases. In this case, $CO_2$ concentration increases at the anion locations, and the anion concentration decreases, even causing oscillations. The difference is that the $CO_2$-anion interaction leads to a decrease in anion concentration, resulting in a monotonous increase in surface charge density. When both the cation-$CO_2$ and anion-CO$_2$ interactions are attractive, as shown in Figure 5e (the default setting), CO$_2$ appears between the cation and anion layers. This increases CO$_2$ concentration and suppresses the densities of cations and anions, leading to a decrease in surface charge density and forming a peak in the volcano plot. For real ILs, the cations and anions exhibit varying degrees of attraction to CO$_2$, leading to a shift in the CO$_2$ density peak between the cation and anion layers at the interface.

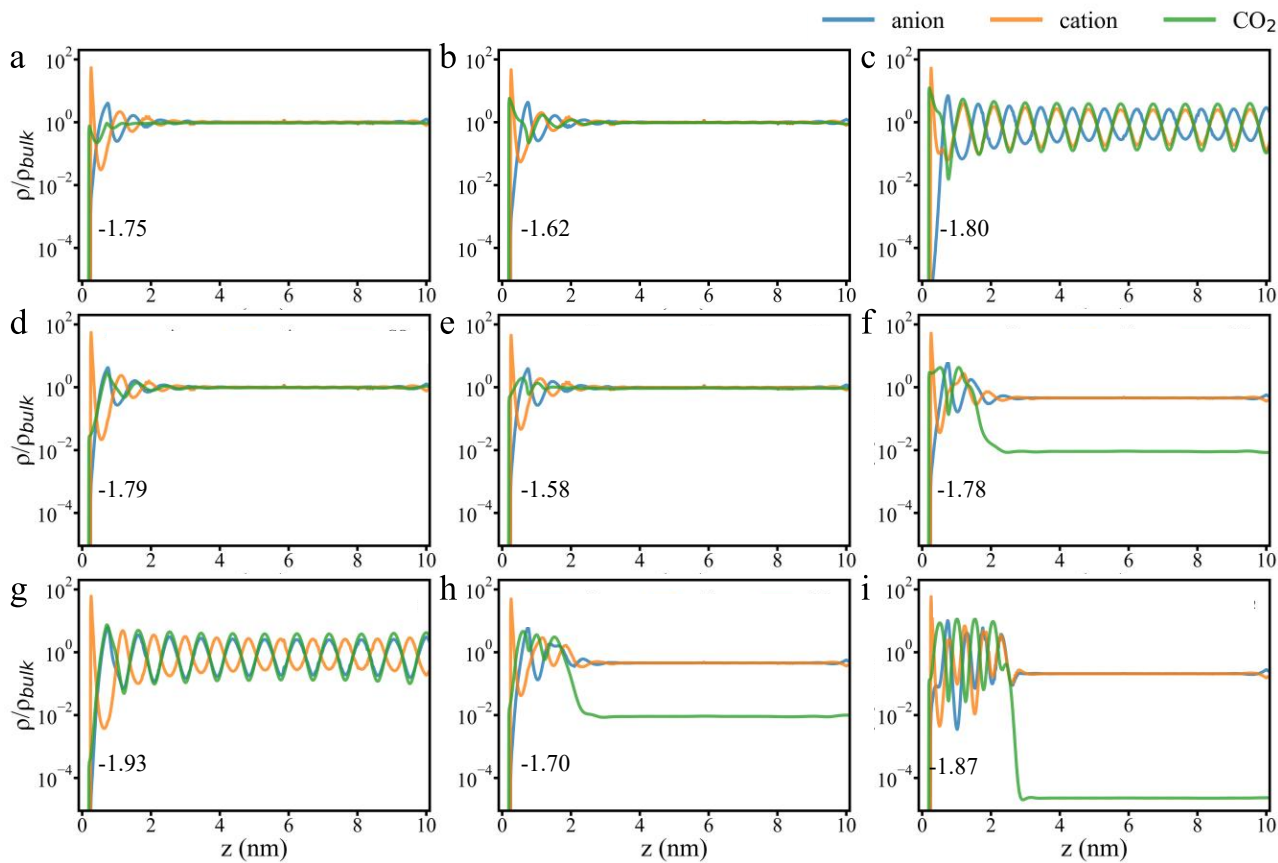

Figure 5. Density profiles under different cation-CO$_2$ ($\varepsilon_{cation\text{-}CO2}$) and anion-CO$_2$ ($\varepsilon_{anion\text{-}CO2}$) interaction energies. At -0.5 V (vs. PZC), the density distributions are shown for the following interaction energy pairs ($\varepsilon_{cation\text{-}CO2}$, $\varepsilon_{anion\text{-}CO2}$): 0.0, 0.0 (a); 5.0, 0.0 (b); 10.0, 0.0 (c); 0.0, 5.0 (d); 5.0, 5.0 (e); 10.0, 5.0 (f); 0.0, 10.0 (g); 5.0, 10.0 (h); and 10.0, 10.0 (i). The unit of energy is k$_B$T. The surface charge densities were marked in each panel.

When both the cations and anions strongly attract CO$_2$, as shown in Figures 5f, 5h, and 5i, we observe a dramatic increase in CO$_2$ concentration near the surface, and the oscillations in the cation and anion densities lead to an anomalous increase in surface charge density. In the bulk phase, long chains of -cation-CO$_2$-anion- are formed, which causes the concentration of all components in the IL to be lower than their bulk densities.

Therefore, it can be concluded that the adsorption effect of ionic liquids on $CO_2$ is not always better with stronger interaction; excessive attractive forces, leading to the formation of long -cation-$CO_2$-anion- chains, hinder $CO_2$ from entering the slit pore. Furthermore, the cations and anions tend to arrange in a layered manner on the electrode surface, and the interaction between $CO_2$ and the cations/anions determines $CO_2$ simultaneous density distribution. It can be speculated that while strong anion-$CO_2$ attraction leads to high $CO_2$ solubility, it also impedes $CO_2$ from contacting the catalyst surface, negatively affecting its conversion activity.

Additionally, Figures S6 and S7 show the changes in the distributions of ionic liquid components on the electrode surface under different interactions between the cations/anions and $CO_2$. It can be observed that the concentrations of cations and anions on the surface show a strong correlation with the surface's attraction or repulsion. While for $\varepsilon_{s-CO2}$, when there is no interaction, the $CO_2$ concentration at the surface is very low, so the repulsive forces have little effect on the surface $CO_2$ concentration, whereas attractive forces significantly increase its concentration at the surface and reduce the value of surface charge density. It is important to note that, as shown in Figures S6c and S6d, when both the electrode interface and the cations/anions exhibit attraction to $CO_2$, two $CO_2$ density peaks appear at the interface—one located at the cation layer and the other between the cation-anion layers. These high-density $CO_2$ peaks result in a very low $CO_2$ density valley between them. Figures S8 and S9 display the distribution changes of IL components under different combinations of cation and anion diameters. Larger ions reduce their concentration and lead to low surface charge density. Notably, the volume differences lead to variations in the accumulation of cations and anions on the surface at 0 V vs PZC, resulting in a net charge difference. This indicates that the PZC has shifted, and additional applied potential is needed to maintain charge neutrality.

**Conclusion**

In this study, we propose a solution-corrected CPM-sol model that combines KS-DFT-based CPM with classical DFT simulations of the electric double layer (EDL), thereby incorporating the effects of the electrode-solution interactions into the CPM model. This CPM-sol model accounts for the electronic accumulation on the electrode surface and the net charge difference caused by the distribution of ions in the solution, reflecting the changes in the Fermi level during the reaction process. Using Cu (100) surface C-C coupling in ionic liquids as an example, this model presents new insights compared to the traditional CPM model. Furthermore, this model can be readily extended to a broad range of electrocatalytic systems, offering improved fidelity in representing the influence of the solution environment.

Additionally, we have systematically explored the influence of electrode surface and ionic liquid interactions on the EDL properties, particularly focusing on how the ionic liquid composition and the interactions between its components affect the electrode surface charge density and the distribution of $CO_2$ near the electrode surface. Our results reveal that the interaction between cations and anions, as well as the interaction between IL components and $CO_2$, plays a pivotal role in determining the surface charge density and, consequently, the electrocatalytic activity. Specifically, repulsive forces between cations and anions lead to a decrease in the local density of the IL components, resulting in a longer oscillation distance to neutralize the effect of the electrode potential. We found that the ionic liquid's ability to adsorb $CO_2$ is not always beneficial to $CO_2$ reduction efficiency. Excessive interaction forces, leading to the formation of long -cation-$CO_2$-anion- chains, hinder $CO_2$ from entering the electrode surface's confined regions. Additionally, the anion-$CO_2$ attraction may impede $CO_2$'s ability to reach the catalytic surface, adversely affecting reaction activity. The distribution of ionic liquid components and their interaction with $CO_2$ significantly affect the electrochemical performance of the electrode in $CO_2$ reduction reactions. By controlling the ionic liquid's composition and its interaction with $CO_2$ and the electrode, we can optimize the catalytic performance and enhance the efficiency of $CO_2$ electroreduction processes.

## Supporting Information

Supporting Information is available and includes detail computational method, Figure S1–S9, and references.

## Conflict of Interest

There is no conflict of interest to report.

## Funding Information

This research is made possible through the financial support from the NSF-DFG Lead Agency Activity in Chemistry and Transport in Confined Spaces under Grant No. NSF 2234013.

TOC Graph

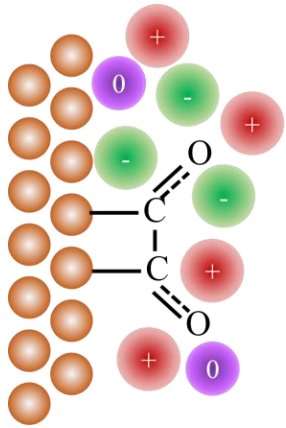 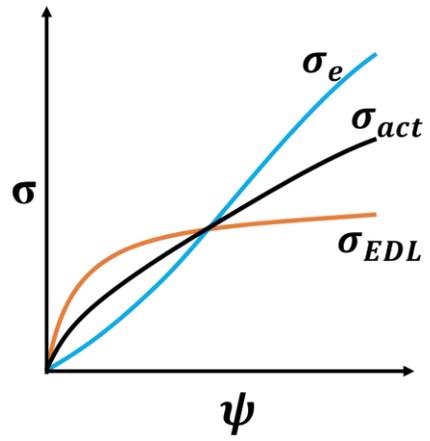